\documentclass[a4paper,fleqn]{cas-dc}
\usepackage{epsfig}
\usepackage{latexsym}
\usepackage{xspace}
\usepackage{hyperref}
\usepackage[utf8]{inputenc}
\usepackage{indentfirst}
\usepackage{enumerate}
\usepackage{listings}
\usepackage{color}

\usepackage[numbers,compress]{natbib}
\newcommand{\sNN}{\sqrt{s_{\rm NN}}}

\newcommand{\eq}[1]{\begin{align} #1 \end{align}}

\usepackage[normalem]{ulem}  % \sout{old text} for strikeout
\usepackage{ulem}
\renewcommand\sout{\bgroup\color{blue} \ULdepth=-.5ex \ULset}

\begin{document}
\let\WriteBookmarks\relax
\def\floatpagepagefraction{1}
\def\textpagefraction{.001}

\shorttitle{Proton cumulants from hydrodynamics and RHIC-BES-II}    
\shortauthors{V. Vovchenko and V. Koch}  
\title [mode = title]{Proton cumulants from hydrodynamics in light of new STAR data}  
\author[1]{Volodymyr Vovchenko}[orcid=0000-0002-2189-4766]
%\cormark[1]
%\fnmark[1]
\ead{vvovchenko@uh.edu}
\affiliation[1]{organization={Department of Physics, University of Houston},
            addressline={3507 Cullen Blvd}, 
            city={Houston, TX},
            postcode={77204}, 
            country={USA}}
            
\author[2]{Volker Koch}[orcid=0000-0002-2157-2791]
%\cormark[2]
%\fnmark[2]
\ead{vkoch@lbl.gov}
\affiliation[2]{organization={Nuclear Science Division, Lawrence Berkeley National Laboratory},
            addressline={1 Cyclotron Rd}, 
            city={Berkeley, CA},
            postcode={94720}, 
            country={USA}}

\begin{abstract}
New measurements of proton number cumulants from the Beam Energy Scan Phase II (BES-II) program at RHIC by the STAR Collaboration provide unprecedented precision and insights into the properties of strongly interacting matter.
This report discusses the measurements in the context of predictions from hydrodynamics, emphasizing the enhanced sensitivity of factorial cumulants and their implications for the search for the QCD critical point.
The experimental data shows enhancement of second-order factorial cumulants and suppression of third-order factorial cumulants relative to the non-critical baseline at $7.7 < \sNN \lesssim 10$~GeV. We discuss implications of this observation for the possible location of the critical point in the QCD phase diagram and opportunities for future measurements of acceptance dependence of factorial cumulants.
\end{abstract}


\maketitle

\section{Introduction}

\begin{figure}[t]
  \centering
  \includegraphics[width=.49\textwidth]{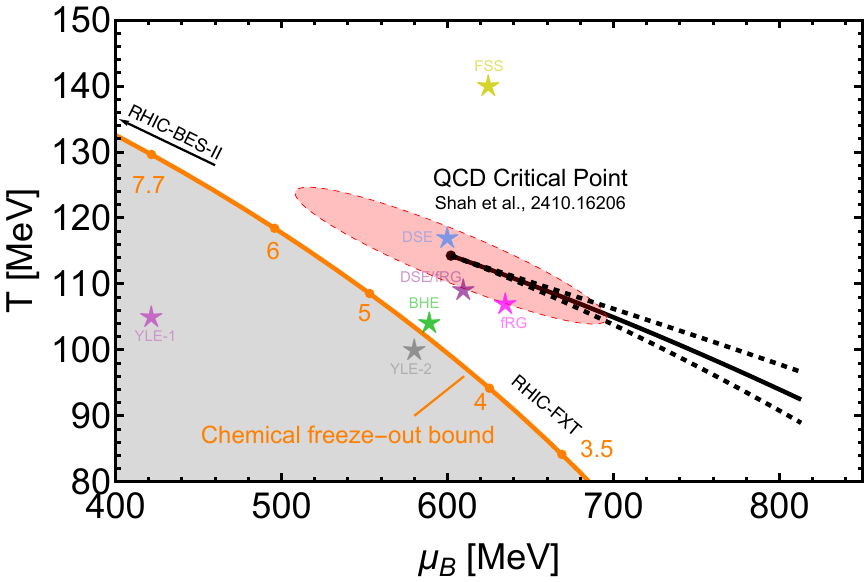}
  \caption{
    Based on~\cite{Shah:2024img}.
    The QCD phase diagram with the CP location estimate from Ref.~\cite{Shah:2024img}~(black point with red covariance ellipse) and other approaches, including Yang-Lee edge singularities~(YLE-1~\cite{Clarke:2024ugt} and YLE-2~\cite{Basar:2023nkp}), functional methods~(fRG~\cite{Fu:2019hdw}, DSE-fRG~\cite{Gao:2020fbl}, DSE~\cite{Gunkel:2021oya}), holography~(BHE~\cite{Hippert:2023bel}), and finite-size scaling (FSS~\cite{Sorensen:2024mry}).
    The orange line represents the chemical freeze-out bound on the CP from Ref.~\cite{Lysenko:2024hqp}, with points on the line corresponding to various collision energies~(in terms of $\sNN$ in GeV).
  }
  \label{fig:CPs}
\end{figure}

Search for the QCD critical point (CP) -- a landmark feature that may exist in the QCD phase diagram -- is a major goal in the studies of QCD under extreme conditions.
This quest has motivated many theoretical studies and is one of the primary goals of the beam energy scan program at RHIC~\cite{STAR:2010vob,Bzdak:2019pkr}.
Determination of the critical point location from first principles is hindered by the fermion sign problem in lattice QCD, which precludes direct simulations at non-zero $\mu_B$.
Recent theoretical efforts are anchored by lattice QCD data at $\mu_B = 0$ and utilize (i) the expected critical behavior of Yang-Lee edge singularities~\cite{Basar:2023nkp,Clarke:2024ugt}, (ii) functional methods~\cite{Fu:2019hdw,Gao:2020fbl,Gunkel:2021oya} and (iii) effective QCD models~\cite{Hippert:2023bel}, or (iv) contours of constant entropy density~\cite{Shah:2024img,Borsanyi:2025dyp}, to estimate the location of the CP at finite $\mu_B$.
These estimates point to a similar range in critical temperature $T_c \sim 100-120$~MeV and chemical potential $\mu_{B,c} \sim 400-650$~MeV.
If accurate, the CP should be accessible by heavy-ion collisions at $\sNN \lesssim 7.7$~GeV~(Fig.~\ref{fig:CPs}).

Heavy-ion collisions scan the QCD phase diagram by varying the collision energy, with lower energies corresponding to higher values of $\mu_B$.
Measurements at different collision energies can constrain QCD CP in a number of ways.
First, the measured hadron abundances are well described by the hadron resonance gas~(HRG) model~\cite{Andronic:2017pug}, which defines the chemical freeze-out line on the phase diagram~\cite{Cleymans:1999st,Cleymans:2005xv,Vovchenko:2015idt}.
The freeze-out line serves as a lower bound in temperature for the CP location at a fixed $\mu_B$, although certain caveats~(e.g. strangeness neutrality) apply~\cite{Lysenko:2024hqp}.

As a more direct CP signature, one utilizes event-by-event fluctuations of hadron numbers~\cite{Stephanov:1999zu}.
Cumulants of baryon number are particularly prominent, as they probe baryochemical potential derivatives of the partition function in equilibrium and can be particularly sensitive to singularities induced by the CP.
In particular, near a CP, the cumulants exhibit universal singular behavior~\cite{Stephanov:1999zu} and reflect the macroscopically large density fluctuations, which lead to the well-known phenomenon of critical opalescence in classical systems.
Of course, the systems created in heavy-ion collisions are too small and short-lived to observe critical opalescence.
Instead, owing to the smallness of the system, one can measure the event-by-event distribution of proton number directly and determine the corresponding cumulants.
Critical fluctuations are not expected to fully equilibrate in heavy-ion collisions, however, model calculations indicate that remnants of the CP signatures are expected to survive~\cite{Stephanov:2008qz,Kuznietsov:2022pcn,Kuznietsov:2024xyn}.
Therefore, if the QCD CP exists in the heavy-ion regime, it should manifest itself in enhanced fluctuations of proton number~\cite{Hatta:2003wn}, in particular, leading to a non-monotonic collision energy dependence of the high-order, non-Gaussian proton number cumulants~\cite{Stephanov:2008qz,Stephanov:2011pb}.

Measurements of proton number cumulants in Au-Au collisions have been performed by the STAR Collaboration at RHIC over the last decade.
Data from beam energy scan~(BES) at $\sNN = 7.7-200$~GeV were published a few years ago~\cite{STAR:2020tga,STAR:2021iop}.
Statistical precision has been improved by over an order of magnitude within BES phase II, with the first preliminary data presented at the CPOD conference in May 2024~\cite{Pandav:CPOD2024}~(preprint now available at~\cite{STAR:2025zdq}), with more data coming from RHIC fixed target program~($\sNN \leq 4.5$~GeV) in the near future.
In this report, we discuss the current theoretical description of proton number cumulants, with a focus on non-critical baseline, and how these predictions compare with the new BES-II data.

\section{Hydrodynamics baseline}

Comparisons of theoretical predictions with experimental data on proton number cumulants require care.
The predictions often assume grand-canonical ensemble in a static volume.
The reality of heavy-ion collisions is different: the system expands into the vacuum, and the measurements are subject to kinematic cuts.
For these reasons, even non-critical contributions to proton number fluctuations, such as global baryon conservation, can make the energy dependence of cumulants non-trivial.
Therefore, a dynamical framework for describing fluctuations in heavy-ion collisions is necessary.
One strategy here is to perform quantitative calculations of the cumulants dependent on the assumed location of the CP, for instance in the framework of hydrodynamics with fluctuations~\cite{Stephanov:2017ghc,An:2021wof,Chattopadhyay:2024jlh}, or hadronic transport~\cite{Sorensen:2020ygf,Kuznietsov:2022pcn}.
The corresponding calculations are not yet on the level of precision needed for a quantitative comparison with the data, but they provide a qualitative understanding of the expected trends.

A different approach utilizes precision calculations of the non-critical baseline, which contains contributions to proton fluctuations that are not related to the CP. 
The most prominent effect is the exact conservation of the global baryon number~\cite{Bzdak:2012an}. 
Baryon conservation suppresses fluctuations in the net proton number, with the magnitude of suppression depending on the size of the acceptance window relative to the size of the system.
The kinematic cuts are the same at all RHIC-BES energies, while the system size is smaller at lower energies.
This leads to a non-flat energy dependence of the cumulants, with a larger suppression at lower energies~\cite{Braun-Munzinger:2020jbk}.
In addition, repulsive baryon interactions, such as due to the hard-core of baryons, can also affect the cumulants~\cite{Vovchenko:2017xad}.

A hydrodynamics-based baseline incorporating baryon conservation and repulsion was developed in Ref.~\cite{Vovchenko:2021kxx}.
It uses (3+1)D relativistic hydrodynamics simulations of Au-Au collisions with MUSIC, tuned to reproduce bulk observables at RHIC-BES energies~\cite{Shen:2020jwv}, and appropriately generalized Cooper-Frye particlization routine~\cite{Vovchenko:2022syc}.
Baryon number conservation was implemented through the subensemble acceptance method, ensuring the canonical treatment of the total (global) baryon number~\cite{Vovchenko:2021yen}.
Repulsion has been implemented through the excluded volume effect, with the repulsion parameter constrained to $b = 1$~fm$^3$ to reproduce lattice QCD susceptibilities~\cite{Vovchenko:2017xad}.

The resulting calculations successfully describe the net-
proton cumulants at $\sNN \gtrsim 20$~GeV.
The cumulants are suppressed by both the baryon conservation and excluded volume, with the suppression factor increasing with decreasing collision energy.
The dominant effect is baryon conservation, but both are needed to describe the BES-I data quantitatively at $\sNN \gtrsim 20$~GeV.
At lower energies~($\sNN \lesssim 20$~GeV), the data show enhancements in second and third-order cumulant ratios that the baseline alone cannot explain.

\section{New STAR data from BES-II and factorial cumulants}

\subsection{Deviations from non-critical baseline and comparison with equilibrium expectations}

Recently, the STAR Collaboration presented new data on proton number cumulants from the BES-II program~\cite{Pandav:CPOD2024}, covering the energy range $\sNN = 7.7-27$~GeV with unprecedented precision.
The new data confirm the trends seen in BES-I for second and third-order cumulant ratios and offer new insights into the behavior of high-order cumulants, such as the indications for a dip in net-proton kurtosis at $\sNN \simeq 20$~GeV.

In addition to ordinary cumulants, factorial cumulants $\hat{C}_n$ have also been presented.\footnote{The STAR Collaboration uses a notation where factorial cumulants are denoted by $\kappa_n$.}
Factorial cumulants are related to ordinary cumulants through a linear relation.
When applied to particle numbers (e.g. proton number instead of net-proton number),
they have the advantage that they probe genuine multi-particle correlations and potentially provide a cleaner signal of critical fluctuations~\cite{Ling:2015yau,Bzdak:2016sxg}.
On the other hand, non-critical contributions~(such as baryon conservation or excluded volume) to high-order factorial cumulants are moderate~\cite{Bzdak:2016jxo,Vovchenko:2017xad}.
Also, the relation between proton and baryon number cumulants~(neglecting antiparticles) is straightforward and transparent on the level of factorial cumulants when all correlations are isospin-blind, $\hat{C}_n^B \sim 2^n \hat{C}_n^p$~\cite{Kitazawa:2012at}.

\begin{figure*}[t]
    \centering
    \includegraphics[width=.80\textwidth]{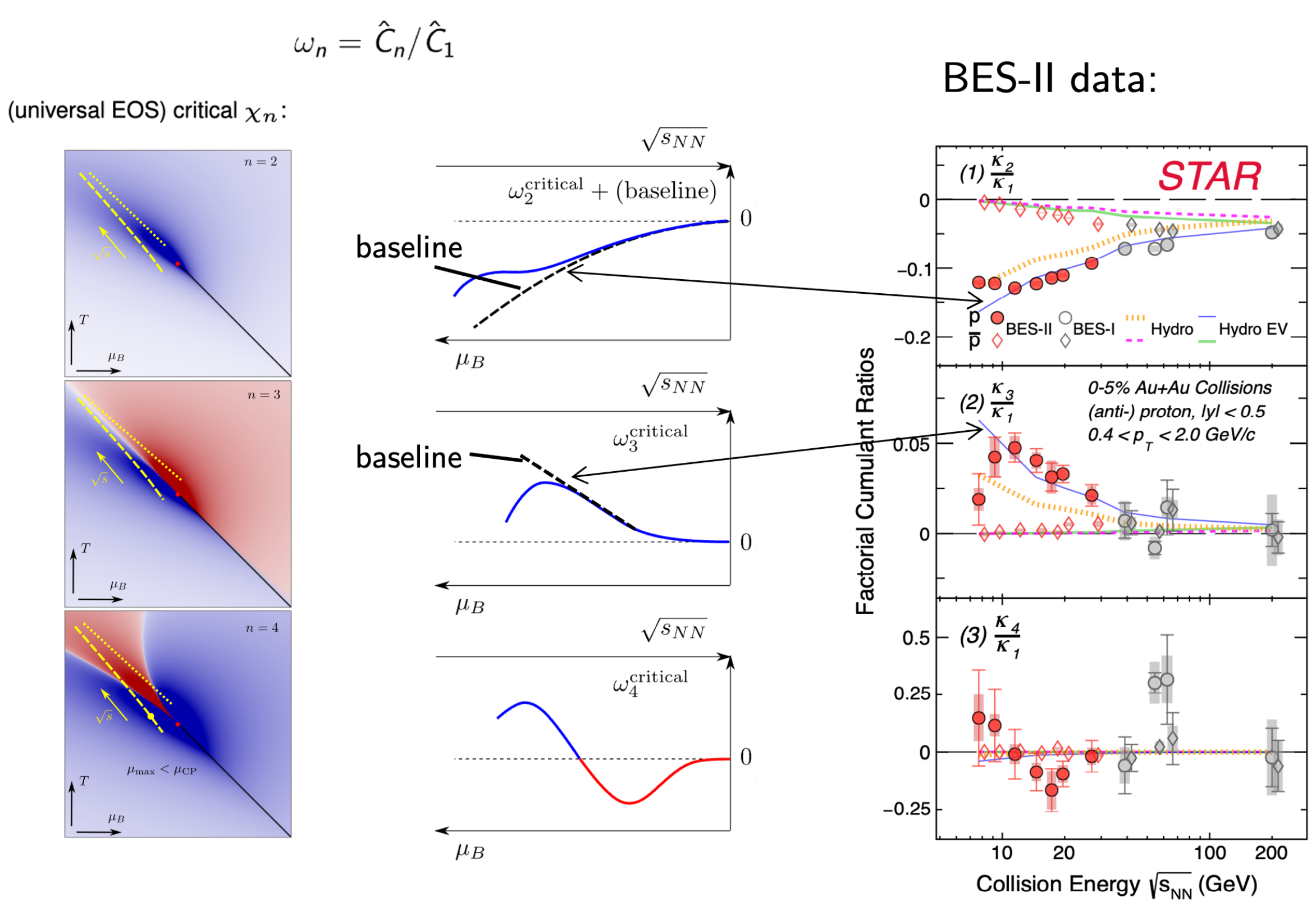}
    \caption{
    Left panel: From Ref.~\cite{Stephanov:2024xkn}. The structure of second-, third- and fourth-order normalized proton factorial cumulants $\omega_n = \hat{C}_n / \hat{C}_1$ near the critical point in the QCD phase diagram and the hypothetical freeze-out trajectory~(dashed line).
    Middle panel: Adapted from~\cite{Stephanov:2024xkn}. Qualitative expectations for the collision energy dependence of proton $\omega_n$ in the presence of the CP. 
    The blue line for $\omega_2$ includes the expected contribution from the non-critical baseline, while the corresponding lines for $\omega_3$ and $\omega_4$ neglect this contribution.
    Right panel: The STAR data on proton and antiproton $\omega_n$ from BES-II~\cite{Pandav:CPOD2024,STAR:2025zdq} compared with the hydrodynamics based non-critical baseline~\cite{Vovchenko:2021kxx}, with~(solid blue line) and without~(dotted orange line) baryon excluded volume effects.
    }
    \label{fig:NewData}
\end{figure*}

The (grand-canonical) equilibrium expectations for the collision energy of the normalized proton factorial cumulants, $\omega_n = \hat{C}_n / \hat{C}_1$ in the presence of the CP are sketched in the middle panel of Fig.~\ref{fig:NewData}, taken from~\cite{Stephanov:2024xkn}.
These are based on the universal structure of factorial cumulants in the vicinity of the CP~\cite{Stephanov:2008qz,Bzdak:2019pkr} and the hypothetical heavy-ion freeze-out trajectory passing near the CP~(left panel of Fig.~\ref{fig:NewData}).
As one traverses the region near the CP by lowering the collision energy, one expects a bump in $\omega_2$, and a non-monotonic behavior in $\omega_3$ and $\omega_4$.
The experimental data on $\omega_2$, shown by the red symbols in the right panel of Fig.~\ref{fig:NewData}, show negative values, $\omega_2 < 0$, throughout, which would naively indicate a violation of the CP expectations.
However, one does observe a non-monotonic behavior and a change of trend around $\sNN = 10$~GeV,
and for meaningful comparisons with the critical expectations, the non-critical baseline must be subtracted.
%can see a change of trend around $\sNN = 10$~GeV.
The data at $\sNN \gtrsim 10$~GeV are consistent with the hydrodynamics baseline~\cite{Vovchenko:2021kxx}, shown in Fig.~\ref{fig:NewData} by the blue line, while the enhancement relative to the baseline is visible at $\sNN \lesssim 10$~GeV, $\omega_2 - \omega_2^{\rm baseline} > 0$. 
As discussed in Ref.~\cite{Stephanov:2024xkn}, this is consistent with the CP expectation.

Measurements of $\omega_3$ show a distinct peak around $\sNN = 10$~GeV, which would, again, be in line with critical expectations.
However, comparing the measurements with the same non-critical baseline as for $\omega_2$, one observes that the latter describes the right side of the peak while failing to capture the left side.
Deviations of the measurements from the baseline are either vanishing~($\sNN \gtrsim 10$~GeV) or negative, $\omega_3 - \omega_3^{\rm baseline} \leq 0$~($7.7 < \sNN \lesssim 10$~GeV).

The data on fourth order factorial cumulant, $\omega_4$, show some indications for a non-monotonic behavior, but the uncertainties are too sizable to draw firm conclusions.

\begin{table}[h]
\centering
\begin{tabular}{|c|c|c|}
\hline
Fact. cumulant & $7.7 < \sNN \lesssim 10$~GeV & $\sNN \gtrsim 10$~GeV \\
\hline
$\omega_2 - \omega_2^{\rm baseline}$ & $>0$ & $\approx 0$ \\
\hline
$\omega_3 - \omega_3^{\rm baseline}$ & $<0$ & $\approx 0$ \\
\hline
\end{tabular}
\caption{Summary of the deviations of second and third-order proton factorial cumulants at RHIC-BES-II~\cite{Pandav:CPOD2024} from the non-critical baseline~\cite{Vovchenko:2021kxx} at different collision energies.}
\label{tab:FCdata}
\end{table}

Table~\ref{tab:FCdata} summarizes the deviations of the proton number factorial cumulants measured by STAR within BES-II from the hydrodynamics-based non-critical baseline of Ref.~\cite{Vovchenko:2021kxx} at different collision energies.

\subsection{Constraining the freeze-out of fluctuations near the critical point}

As discussed above, the experimental data show clear deviations from non-critical baseline at $7.7 < \sNN \lesssim 10$~GeV.
The exact reason for these deviations remains to be established.
For instance, it is possible that additional non-critical effects not captured by the baseline, such as for example volume fluctuations or baryon stopping, play an important role here.
Alternatively, the deviations could be due to the presence of critical fluctuations.
Let us consider here the latter possibility.

Assuming the observed deviations do stem from critical behavior, one may put constraints on where in the QCD phase diagram the proton number fluctuations freeze out.
The corresponding methodology has been developed in Ref.~\cite{Bzdak:2016sxg}, where a 3D-Ising and the preliminary~(at that time) RHIC-BES-I data were used in the analysis.
The signs of $\hat{C}_2$ and $\hat{C}_3$ relative to the baseline provide qualitative information on where the fluctuations may freeze out in the vicinity of the CP.
This analysis is repeated here, with a couple of modifications.
First, we use the new STAR data from BES-II, and we subtract the non-critical baseline from the data before applying the analysis.
Second, we use the van der Waals model for nuclear matter to depict the phase diagram in the familiar $T$ and $\mu_B$ coordinates instead of the Ising variables.\footnote{The analysis could, in principle, be repeated with the 3D-Ising model with an appropriate mapping of the Ising variables onto $T$ and $\mu_B$ plane~\cite{Parotto:2018pwx}.}

\begin{figure}[t]
    \centering
    \includegraphics[width=.49\textwidth]{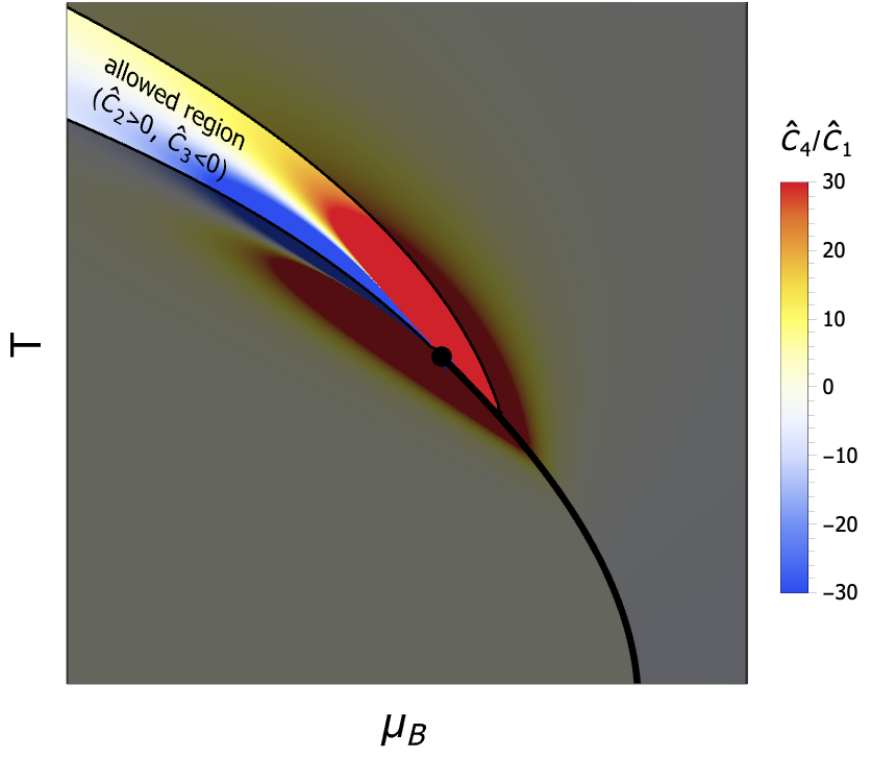}
    \caption{
    Exclusion plot for the freeze-out of proton number fluctuations near the CP based on the sign of the second-order factorial cumulant $\hat{C}_2$ and the third-order factorial cumulant $\hat{C}_3$.
    The unshaded region~(the narrow band labeled "$\hat{C}_2 > 0$, $\hat{C}_3 < 0$") represents the allowed region for the freeze-out of fluctuations given by $\hat{C}_2 > 0$ and $\hat{C}_3 < 0$, which qualilatively agrees with the STAR data corrected for non-critical baseline.
    The density plot represents the expected behavior of $\omega_4 = \hat{C}_4/\hat{C}_1$. 
    The analysis of critical behavior is based on the van der Waals model for nuclear matter.
    }
    \label{fig:CPexcl}
\end{figure}

Figure~\ref{fig:CPexcl} shows the result of the analysis.
In the shaded region either $\hat{C}_2 < 0$ or $\hat{C}_3 > 0$~(or both), which is inconsistent with the deviations of the STAR data from the baseline.
The unshaded region corresponds to $\hat{C}_2 > 0$ and $\hat{C}_3 < 0$, which is in qualitative agreement with the STAR data.
One observes that the allowed region mainly corresponds to the QGP side of the transition near the crossover region.
This observation would indicate that proton fluctuations may freeze out earlier than when the system reaches the conventional chemical freeze-out, indicating out-of-equilibrium dynamics. 
Such an early freeze-out of fluctuations could reflect on-equilibrium effects and critical slowing down, as large correlation lengths require time to develop~\cite{Berdnikov:1999ph,Mukherjee:2015swa}.

\subsection{Nuclear liquid-gas transition}

\begin{figure}[t]
    \centering
    \includegraphics[width=.49\textwidth]{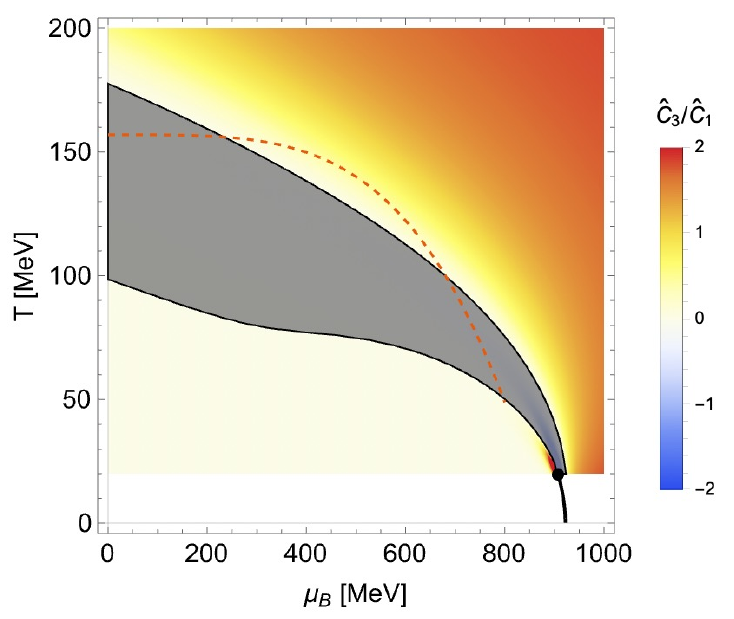}
    \caption{
    Third-order factorial cumulant ratio $\omega_3 = C_3/C_1$~(grand-canonical limit) as a function of $T$ and $\mu_B$ in a van der Waals-like HRG model containing nuclear liquid-gas transition~\cite{Vovchenko:2017ygz}. 
    The dashed line represents the chemical freeze-out line in a non-ideal HRG model~\cite{Poberezhnyuk:2019pxs}.
    The shaded region corresponds to $\hat{C}_3 < 0$.
    }
    \label{fig:c3-LGHRG}
\end{figure}

Nuclear liquid-gas transition takes place at low temperatures and high baryon densities, with a CP estimated to be at $T_c \sim 20$~MeV and $\mu_B \sim 900$~MeV~\cite{Elliott:2013pna}.
Although at first glance this may seem unrelated to the QCD CP and heavy-ion collisions, model calculations suggest that remnants of the nuclear CP extend to higher temperatures and lower baryon densities in high-order cumulants of baryon number~\cite{Mukherjee:2016nhb,Vovchenko:2016rkn,Sorensen:2020ygf}, in particular along the chemical freeze-out line~\cite{Vovchenko:2017ayq}.
The same holds also for factorial cumulants.
Figure~\ref{fig:c3-LGHRG} shows the third-order factorial cumulant $\omega_3 = C_3/C_1$ on the QCD phase diagram calculated in a HRG model with van der Waals-like interactions~\cite{Vovchenko:2017ygz}, along with the chemical freeze-out line in that model~\cite{Poberezhnyuk:2019pxs}.
Calculations are performed in the grand-canonical limit and are not yet appropriate for quantitative comparisons with data.
Nevertheless, they show that third-order cumulant may exhibit a non-monotonic behavior along the freeze-out line solely due to the presence of the nuclear liquid-gas transition.

With the new data coming from RHIC fixed target programme at $\sNN = 3-4.5$~GeV, as well as the upcoming CBM experiment at FAIR, the relevance of nuclear liquid-gas transition for the interpretation of proton number cumulants in heavy-ion collisions will become more pronounced.
In addition, at low energies, a significant fraction of the baryon number is carried by light nuclei, such as deuterons and tritons, and their inclusion in the analysis of proton number cumulants is necessary.

\subsection{Acceptance dependence and long-range correlations}

Factorial cumulants are also useful probes of long-range correlations in heavy-ion collisions.
In particular, the reduced factorial cumulants, defined as $\hat{c}_n = \hat{C}_n / \hat{C}_1^n$ for both protons and antiprotons, can distinguish between long-range correlations and short-range correlations~\cite{Bzdak:2017ltv}.
If long-range correlations are the only source of correlations, then $\hat{c}_n$ should be independent of the acceptance window:
\eq{\label{eq:scscale}
c_n = \hat{C}_n / \hat{C}_1^n = \rm const.
}
Sources of long-range correlations include global baryon conservation, volume fluctuations, and a combination of both.
However, if short-range correlations are present, for instance, due to critical fluctuations or local charge conservation, then $\hat{c}_n$ will depend on the acceptance window and violate the scaling in Eq.~\eqref{eq:scscale}.
Therefore, a detailed analysis of the rapidity and/or $p_T$ acceptance dependence of factorial cumulants at each collision energy can provide insights into the nature of correlations in heavy-ion collisions.
Notably, since the presence of volume fluctuations preserves the scaling in Eq.~\eqref{eq:scscale}, the corresponding analysis can be performed without any corrections for volume fluctuations such as CBWC.

The scaled factorial cumulants $\hat{c}_n$ can be reconstructed from the available measurements of factorial cumulants $\hat{C}_n$ at BES-I~\cite{STAR:2021iop}.
Acceptance behavior of $\hat{c}_2$ from BES-I does indicate that the scaling in Eq.~\eqref{eq:scscale} is largely satisfied for both protons and antiprotons.
However, it also reveals a significant difference between protons and antiprotons, the latter showing stronger reduced anticorrelations, which are not described by the hydrodynamics baseline~\cite{Bzdak:2025rhp}.
The explanation for this behavior of antiproton factorial cumulants is not yet clear, the prelimary BES-II data on $\hat{C}_2^{\bar p}/\hat{C}_1^{\bar p}$~(Fig.~\ref{fig:NewData}) similarly indicate a significant deviation from the hydrodynamics baseline.

A meaningful analysis of high-order $\hat{c}_3$ and $\hat{c}_4$ is not yet possible due to the limited statistics of the BES-I data.
The precision BES-II data will allow one to perform a detailed analysis of the acceptance dependence high-order reduced factorial cumulants of protons and antiprotons and shed light on the nature of correlations in heavy-ion collisions.
In particular, the presence of the CP should lead to a violation of the long-range correlation scaling~\eqref{eq:scscale}, especially in high-order factorial cumulants. 
The analysis of acceptance dependence thus complements that of collision energy dependence discussed and presented above.

\section{Summary}

New measurements from the RHIC-BES-II program at RHIC provide unprecedented precision and insights into the behavior of proton number cumulants in heavy-ion collisions.
Prediction from hydrodynamics with baryon conservation and excluded volume effects provide a non-critical baseline for the analysis of the data, which is in excellent agreement with the data at $\sNN \gtrsim 20$~GeV. 
This observation alone strongly disfavors the existence of the CP at small baryon densities, $\mu_B \lesssim 200$~MeV.

Further insight is gained through the analysis of factorial cumulants, which are sensitive to genuine multi-particle correlations and can provide cleaner signals of critical fluctuations.
Pronounced deviations from the baseline are observed at $7.7 \sNN \lesssim 10$~GeV, with the second- and the third-order factorial cumulants showing enhancement and suppression, respectively, relative to the baseline.
It remains to be seen whether these deviations are due to critical fluctuations or other non-critical effects, such as volume fluctuations or baryon stopping.
If one does attribute the deviations to critical fluctuations, this could imply that the freeze-out of proton number fluctuations at $7.7 < \sNN \lesssim 10$~GeV takes place on the crossover/QGP side of the CP, indicating that CP might exist at collision energies $\sNN \lesssim 7.7$~GeV, or $\mu_B \gtrsim 400$~MeV.
Such a conclusion would be in line with the available theoretical estimates of the CP location in the QCD phase diagram anchored with lattice QCD constraints.

The analysis of collision energy dependence of factorial cumulants should be complemented by the analysis of acceptance dependence.
In particular, critical behavior is expected to induce (high-order) local correlations, which would violate the long-range correlation scaling in Eq.~\eqref{eq:scscale} for scaled factorial cumulants.
This can be tested experimentally even without volume fluctuation corrections, such as CBWC.

Data at $\sNN < 7.7$~GeV from the RHIC fixed target program and later at FAIR, will provide additional insights into the nature of correlations in heavy-ion collisions and their possible link to the CP.
Improved treatment of baryon stopping, centrality selection, and effects of nuclear liquid-gas transition and light nuclei will be necessary for the analysis of the data at low energies.

\section*{Acknowledgments}
The authors thank Ashish Pandav and the STAR Collaboration for providing the comparison of the preliminary data on proton number cumulants from BES-II with the hydrodynamics based baseline.
The authors thank Adam Bzdak for the fruitful discussions and collaboration in analyzing the acceptance dependence of factorial cumulants.
This material is based upon work supported
by the U.S. Department of Energy, Office of Science, Office of Nuclear
Physics, under contract number DE-AC02-05CH11231.

\bibliography{main}

\end{document}